# Software-defined optical networking applications enabled by programmable integrated photonics

Zhenyun Xie,[1, *] David Sánchez-Jácome,[1] Luis Torrijos-Morán[1] And Daniel Pérez-López[1, *]

[1]iPronics Programmable Photonics S.L., Av. de Blasco Ibáñez, 25, entresuelo, 46010 Valencia
*Corresponding author: zhenyun.xie@ipronics.com , daniel.perez@ipronics.com

**Data center networks are experiencing unprecedented exponential growth, mostly driven by the continuous computing demands in machine learning and artificial intelligence algorithms. Within this realm, optical networking offers numerous advantages, including low latency, energy efficiency, and bandwidth transparency, positioning it as a compelling alternative to its electronic counterparts. In this work, we showcase a range of software-defined optical networking applications deployed on a general-purpose programmable integrated photonic processor. Leveraging graph-based theory, we experimentally demonstrate dynamic optical interconnects, circuit switching, and multicasting on the same photonic platform, yielding remarkable results in terms of crosstalk and reconfiguration speed. Our approach harnesses the benefits of reconfigurability and reliability, paving the way for a new generation of high-performance optical devices tailored for data center and computing clusters.**

## 1. INTRODUCTION

The ever-growing demand for more powerful machine learning (ML) models is producing a rapid increase in the amount of traffic within datacom networks[1], [2]. To cope with these requirements, data centers (DC) are evolving towards disaggregated architectures and simultaneously increasing the bandwidth of the optical links up to hundreds of gigabits per second. Moreover, it is expected to go beyond terabits per second in the foreseeable future[3], [4]. However, electrical switches employed in every stage of the network are facing fundamental limitations in terms of energy consumption and bandwidth to keep the pace of ML workloads[5], [6], [7]. Optical point-to-point interconnects and electronic switches have been widely used in DC for decades, and applying optics to exploit also circuit switching is becoming a viable alternative[8]. Optical circuit switches (OCS) are intended to keep up with the bandwidth scaling as they are transparent to data rates, agnostic to modulation formats, energy efficient and contribute to reduced latency[9]. Some OCS solutions based on electro-mechanical systems (MEMS) micro-mirrors have already been deployed in DC networks, presenting substantial improvements in terms of performance and radix scalability[10], [11]. Despite the advantages of MEMS-based OCS, reliability, form factor and the lack of processing functions are still under study.

Photonic integrated circuits (PICs) have attracted increasing attention for datacom due to the possibility of building complex optical processors that could serve as OCS[12], [13], [14], [15]. A substantial proportion of these circuits belong to the category of application-specific photonic integrated circuits (ASPICs). However, the development time from the initial design to the final fabrication involves multiple iterative cycles, which leads to long time-to-market intervals and also high-cost. As an alternative, general-purpose programmable integrated photonics (PIP) enable software-defined circuit reconfiguration, with more versatile functionalities and enhanced adaptability[16], [17]. The idea of PIP stems from using programmable waveguide meshes to implement different hardware configurations with a large variety of applications such as radiofrequency processing, beamforming, optical filtering, and computing, among others. One of the most predominant configurations in PIPs is the hexagonal mesh, which enables the recirculation of light to synthesize different photonic circuits[18], [19]. Here, the programmable unit cell (PUC), based on 2x2 Mach-Zehnder interferometers (MZI), can be reconfigured into three distinctive operational states: bar, cross, and tunable coupler[20], [21]. By properly manipulating these PUCs, a dynamic interplay of several optical circuits, such as optical interconnects, splitters, and switches can be programmed. The conventional control of these programmable circuits consists of tuning each individual PUC sequentially to obtain the desired behavior. This method requires human intervention, resulting in low efficiency and potential errors. Consequently, as the waveguide mesh's integration level scales, the mesh validation and calibration procedures' complexity is accentuated, underscoring the need for an automatic control plane provided through software stack[22].

Lately, the development of routing algorithms based on graph theory to program PIP circuits is gaining popularity among the scientific community[23], [24], [25], [26]. As a mechanism to control the path of light propagation, the optical ports are modeled as graph vertices, and performance metrics for each connection are incorporated into the edge attributes. Tunable couplers can be modelled either using a conventional representation or using dummy and artificial nodes[26], [27]. In this regard, path routing algorithms and delay lines have been explored to develop point-to-point optical interconnects in a hexagonal mesh, with fault-tolerant and self-healing capabilities but applied in a short-scale circuit of a few PUCs[23]. More recently, a novel N x N automated switching algorithm has also been reported theoretically, which computes a mesh configuration consisting of up to hundreds of unit cells without conflict[24]. Nevertheless, experimental demonstration of switching algorithms based on graph theory is still to be developed, as well as more complex functionalities such as 1 x N splitting algorithms for multicasting purposes in networking applications. Besides, there is still room for improvement in enhancing the performance of interconnect algorithms, particularly when efficiently identifying interconnect paths in larger mesh sizes.

In this work, we provide a comprehensive implementation of multiple software-defined on-chip optical networking functions based on graph theory. We experimentally demonstrate several applications such as optical interconnect, circuit switching, and novel multicasting algorithms running on a single PIP platform with fast reconfiguration times and high-performance specs. A software layer is also presented to program the PIC, which facilitates the control of the hardware. Finally, we compare the results with similar works and discuss the impact of our work, as well as comment the further research and perspectives.

## 2. ON-CHIP OPTICAL NETWORKING PROCESSOR

The programmable processor consists of a hexagonal arrangement of 72 Programmable Unit Cells (PUCs) and 28 optical ports connected to photodetectors[28]. The schematic of the core is shown in Fig.(a). It should be noted that bottom ports cannot be used as optical I/O they are not directly connected to photodetectors in the hardware design. The PUCs can be configured to the bar, cross and tunable coupler states by properly adjusting the phase shifters within the 2x2 MZI. Subsequently, by suitable programming of all the PUCs inside the hexagonal mesh, the optical processor can be configured to provide several applications and functions.

To control the photonic processor, a software layer based on graphs is used. Graph theory provides a framework for analyzing and solving problems related to connectivity, paths, cycles, and other properties of graphs. Fig. 1 (b) shows the graph representing the hexagonal mesh. Here, all the connectivity and topology of hexagonal mesh can be abstracted into a graph, in which the optical ports of the PUCs are represented as nodes in the graph, and the flow of light in the waveguides can then represented by edges connecting these nodes. One challenge in graph representation of hexagonal mesh is the feedback loops in the circuit. To address this, we chose the directed graph with eight artificial nodes[27] for our model, where each port of PUC is represented with two artificial nodes, for inbound and outbound of the signal. The PUC phase shifters are modeled by arcs with the direction of input and output ports.

The graph representation incorporates performance metrics for each connection, which means each arc is assigned a weight as expressed by the following equation.

$$w = c_1 \cdot IL + c_2 \cdot BUL + C_3 \cdot P_c + \ldots \quad (1)$$

where IL is the PUC insertion loss, BUL is the PUC basic unit length, and $P_c$ is the PUC power consumption. Hence the weight can be calculated according to these Figure of Merit (FoM) by only changing the distribution of weights $\{c_i\}$ [23]. The graph-based mesh can be systematically connected in a predetermined manner, thus enabling to simulate a symmetrical hexagonal photonic mesh of different sizes.

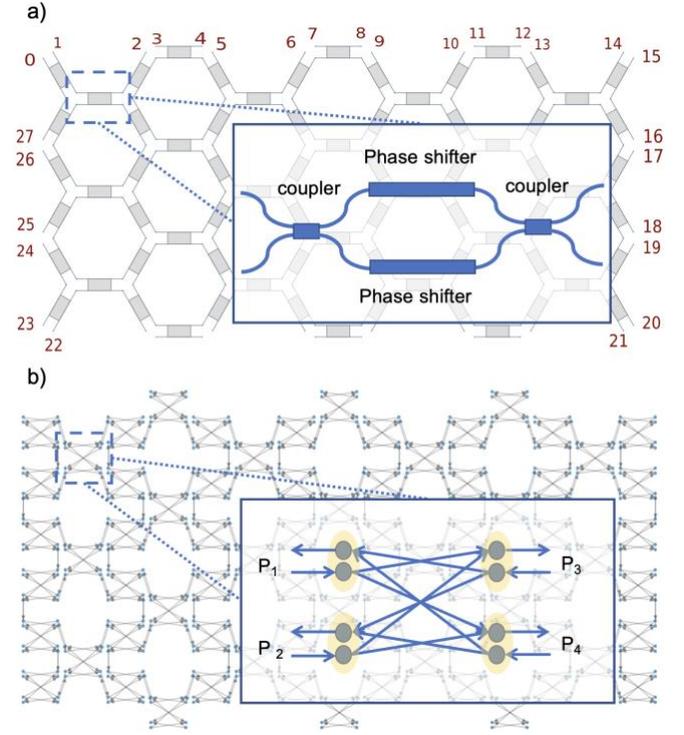

Fig. 1 (a) Schematic representation of the programmable processor core based on hexagonal cells. The inset shows a PUC consisting of a 2x2 MZI with phase shifters at each arm. (b) Graph-based representation of the photonic mesh. The inset depicts a detailed representation of the edges and vertices in a PUC.

## 3. RESULTS

### 3.1 Optical interconnect (1x1)

Reconfigurable optical interconnects are of great interest in optical networks as they enable flexible connections of terabit data rates with high-performance operation[3], [4]. These interconnects demand reduced latency and energy efficiency between input and output (I/O) ports, and therefore minimal insertion loss and shortest distance is highly desired. The software layer demonstrated herein provides an interconnect algorithm based on Dijkstra[29], [30], which can automatically provide an optimum path configuration. This application can also provide self-healing capabilities to allocate alternative optical paths with lower insertion loss when needed.

## 3.2 Optical circuit switching (NxN)

Optical circuit switching is a technology that allows to configure the physical layer of an optical network through software instructions. It holds the potential to improve the cost, latency and power consumption used by ML/AI workloads within data centers. When the traffic circulating in a network is deterministic the physical layer can be used to route the data directly and the need for an Electronic Packet Switch (EPS) can be avoided. In this section, we present the implementation of an on-chip software-defined switch that enables the dynamic control of optical waveguide paths for multiple input and output channels. Such a system has been synthesized on the hexagonal core introduced previously and has been designed to cross-connect 6 pairs of inputs-outputs. To enable the automatic reconfigurability of the full 6-by-6 switch matrix we have adapted three algorithms and implemented them in our software stack. The algorithms explored in this work are sequential routing [23], graph-based with edge weight penalty [24], and analytical decomposition [31]. The first two are more general as they make use of iterative algorithms to find the non-conflicting routes between the set of port pairs. The first is the most time consuming of the three as it relies on testing all possible I/O port routes until it finds a non-conflicting solution. The edge penalty approach is an iterative algorithm that makes use of dynamic weight updates in the graph to penalize conflicting path sections. Once the weight is updated the routine will attempt to find a new route in the next iteration. The *max_iter* parameter (see Table 1) regulates the maximum number of iterations for paths re-routing in instances of edge conflicts, when no routing solution exists for certain input/output port combinations. The third approach is fully analytical and relies on the mathematical decomposition of the switch matrix to a Spanke-Benes network synthesized on the hexagonal mesh. The solution in this case is immediate due to the analytical nature of the problem, however, the set of hexagonal core ports that can be used are restricted by the feed-forward network architecture. On the matter of switching speed, the reconfiguration time of the device will be dominated by the thermo-optic actuator speed, in the order of microseconds, compared to hundreds of milliseconds in the case of MEMS-based approaches [10], [11].

The first and third approaches are straightforward to implement and have been previously documented [23], [31]. Here we document in deeper extent the second approach that implements the edge penalty algorithm since it is more efficient than the brute force algorithm (first approach) and is not constrained to the ports of a Spanke-Benes network (third approach). Refer to Table 1 for a pseudo-code of the algorithm. In lines 1 and 2, the best paths for each I/O port pair are identified. The next step is to check whether these paths contain conflict edges (line 4), as conflicting edges within the identified paths will be penalized with a higher weight (line 6). What we define as a conflicting edge is a PUC that belongs to more than two paths and has a different state (cross/bar) in at least two of these paths. A switching network needs to have zero conflicting edges to work. Once the problematic edges have been penalized, we recompute new optimal paths for the I/O ports pairs considering the new edge weights and updating the preprocessing paths configuration list (line 7, 8). If no conflicts are detected (lines 10, 11), the configuration of optimal paths will be saved and the loop will be stopped, otherwise the iterative process of weight update will continue until a zero-conflict configuration is found or the max number of iterations is reached. If the number of iterations goes beyond the *max_iter* value, the problem will be considered unsolved. Hence, this variable represents the trade-off between the probability of finding a solution and the time spent searching for a zero-conflict configuration.

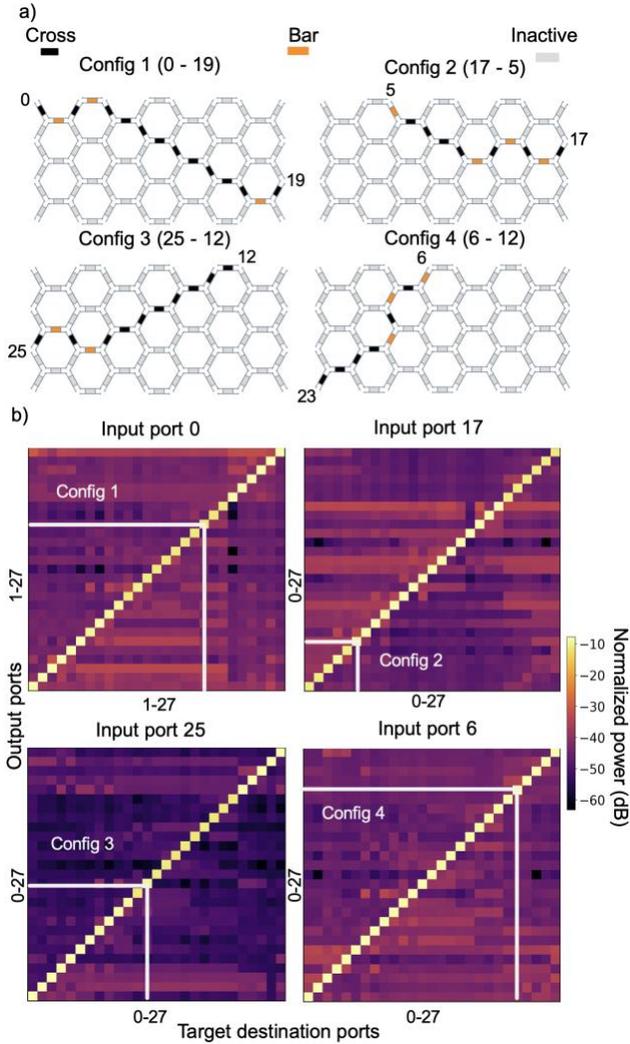

Fig. 2. Optical interconnect experiment. (a) Programmable processor configurations considering different input ports. (b) Colormap of the normalized optical power measured at output ports for different interconnect configurations. The white lines indicate the configuration shown in the schematic of the programmable processor.

The results of the optical interconnect experiments at a central wavelength of 1550 nm and the processor configuration are shown in Fig. 2. Ports 0, 37, 17, 6 are used as input ports, and up to 27 interconnects to other available output ports are programmed. As it can be seen in Fig 2 (b), all the paths are configured successfully, showing diagonal lines in the colormap, each one corresponding to a given interconnect. The power of the measured output is normalized to the input power of the laser. The insertion loss at each target port is between 7.7 dB and 10.5 dB, depending on the number of PUCs used to configure the path. In the experiment, the longest interconnect path consists of 15 PUCs, while the shortest is only 2 PUCs which produces a certain power imbalance between outputs. Furthermore, optical power is also measured at the rest of the non-targeted output ports, with an average leakage of -30 dB. The algorithm allows the simultaneous identification of up to 400 possible paths with an average time of about 0.4 microseconds per path.

**Table 1. Pseudo-code of switch application**

| | Algorithm: auto_switch |
|---|---|
| | **Function auto_switch** *(i_o_ports, max_iter, algorithm = "edge_penalty")* |
| 1 | *preprocessing_paths* ← ROUTING_PREPROCESSING *(i_o_ports)*<br># Obtain all optimum interconnects of target I/O pairs |
| 2 | *best_paths* ← ∅ |
| 3 | **while** iter < max_iter: |
| 4 |    *conflit_edges, conflict_ports* ← GET_CONFLICT_EDGE *(preprocessing_paths)* |
| 5 |    **if** *conflict_edges*: |
| 6 |       **penalize** *conflict_edges* |
| 7 |       **get** *new_paths* # The best paths of conflict_ports |
| 8 |       **update** *preprocessing_paths* **with** *new_paths* |
| 9 |    **else:** |
| 10 |       *best_paths* ← GET_BEST_PATHS *(preprocessing_paths, best_paths)* |
| 11 | **if** *best_paths* **is** ∅: |
| 12 |    **raise** *exception* |
| 13 | **return** *best_paths* |

In Fig. 3 we present the implementation and the measurements at a central wavelength of 1550 nm taken from a 6x6 optical circuit switch. In (a) notice that we have selected six ports from the left side of the hexagonal mesh [0, 27, 26, 25, 24, 23] as inputs and other six ports from the right side [15, 16, 17, 18, 19, 20] as outputs, the selected ports from both sides are named from 1- 6 in blue color and red color respectively. To measure the performance of the network, six distinct switch configurations have been implemented using a laser source with 5 dBm input power connected to port 1. The switch was then software defined to synthesize 6 different configurations between this input and all the possible outputs. Once the network was defined, the power of each output port was measured and normalized with the input source, resulting in the characterization of 6 switch matrices. The process was repeated for the other 5 inputs leading to the characterization of 36 switch matrices in total. Fig. 3 (a) illustrates an example of a 6x6 switch circuit configured on a programmable processor with all PUCs activated. In Fig. 3 (b), each column in the subfigure represents the optical power monitored in a specific output port, and the different marker shapes correspond to the 6 different input ports. Our results demonstrate the optical circuit switch's performance showcasing accurate switching of multiple inputs to their respective target outputs. Notably, the distances of all switch circuit paths are consistent, passing through 15 PUCs. As a result, we observed -10 dB (± 0.5) insertion losses in all 6 output ports for each configuration. Optical crosstalk is determined by checking the difference between the optical power from the specified input output pair and the highest leakage power from other input ports to the same output port. This distinction is visually represented by the distance between two blue lines in Fig. 3(b), with a typical value of -25 dB. This metric varies based on the switch configuration, the best scenario of crosstalk observed in our experiments is 26 dB for the switch configuration: [1:4, 2:5, 3:6, 4:1, 5:2, 6:3]. The source of crosstalk is mainly from PUC performance, as well as small drifts due to auto-calibration and auto-characterization.

Using the same set of selected I/O ports, we evaluate the feasibility of the algorithm. For the 6x6 circuit switch, there are a total of 6! possible combinations, representing distinct switch circuits. The algorithm demonstrated its reliability by generating feasible solutions for all 720 cases without any observed instances of failure. This fact makes the software layer developed for this work a key differentiator with respect to previous manually implemented solutions.

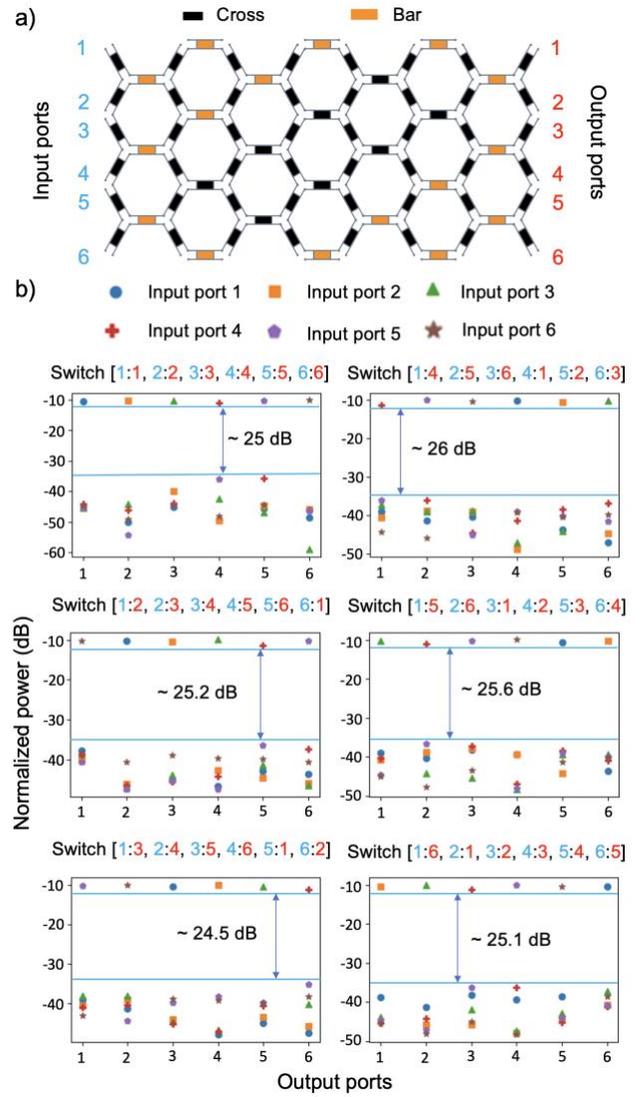

Fig. 3. Optical circuit switching experiment. (a) Programmable processor 6x6 circuit switch configuration. Input ports 0, 39, 38, 37, 36, 35 and output ports 15, 16, 17, 18, 19, 20 are numbered from 1 – 6 each in blue color and red color respectively. (b) 6x6 switch matrix for 6 different configurations. Each subplot demonstrates the output of a specific switch configuration, with the X-axis depicting the output ports in the mesh and the Y-axis representing the measured optical power levels at the output ports. The six input ports are visually represented by six distinct markers in the figure.

### 3.3 Optical multicasting (1xN)

Optical multicasting allows for a single optical signal to be efficiently distributed across multiple receiving nodes or devices simultaneously. This is particularly advantageous in scenarios where the same data needs to be delivered to several destinations within a DC or set of clusters, simultaneously. This functionality is currently not available in the physical layer of most networks so if deployed under the right software stack support it has the potential to open the door to new, cost efficient and disruptive applications. One added advantage brought in by PIP is the software defined control of the splitting ratio across the outgoing signals powered by the fine tuning of driving phases applied to the PUCs. As a consequence of this, the operator can then control how much power is allocated to different receiving nodes in a dynamic manner.

In this work, we have designed an auto-multicast algorithm that can accommodate any number of desired outputs from a given input. It determines the optimum path to each output, and efficiently auto-configures the splitting PUC with a ratio that ensures every output port receives the same power, for a traditional multicast operation, or the specified power ratio, under custom operation. To achieve this, the auto-multicast algorithm can be translated into a single-source shortest path problem represented by a directed acyclic graph (directed tree)[32]. A simple example of a 1-by-3 network is demonstrated in Fig. 4(a). The graph representation of the multicast network is useful when determining how the multicasting circuit can be mapped on the hexagonal mesh as presented in Fig. 4(a). Building upon this intuitive target, our proposed algorithm implements the shortest-path tree circuit based on the pseudo-code in Table 2. In the latter, the initialization of variables starts in lines 1 – 3 and the optimal interconnect paths to each output port will be found and saved in the set *PATHS* in line 4. The Tunable Couplers (TCs) PUCs can be selected from the *PATHS* set under the condition that TCs are those PUCs present in more than two paths with differing states between them (line 5), as denoted by:

$$TC = \{PUC \in P_{i,j} \,\&\, S_i \neq S_j \,\forall\, i \neq j, \, i,j \leq N, \, i,j \in \mathbb{Z}\} \quad (2)$$

where P is the set of paths to each output port, $S_i$ and $S_j$ are the state of a PUC in paths *i* and *j*, respectively. *I* & *j* both are integers and lower or equal than N, the number of outputs. To determine the splitting ratio "*k*" for these TCs, we initiate the calculation from the child nodes to the parent nodes and all the way up to the root (line 6). This approach facilitates calculating the power distribution across each output port (line 16). When light is divided at a TC it will traverse two different paths, defined here as cross and bar paths, until it reaches the output ports. The inherent insertion loss difference between these paths will cause an undesired power imbalance at the outputs, i.e., the power ratio between the outputs will not be as intended. Therefore, it is vital to account for the insertion losses in both paths (cross and bar) coming out of a TC and compensate them to achieve the targeted power distribution, see Eq. (3)

$$k = \frac{k_T}{(1-k_T) \times 10^{\frac{IL_b - IL_c}{10}} + k_T} \quad (3)$$

where "$k_T$" is the ideal splitting ratio of the tunable coupler PUC, and $IL_b$ and $IL_c$ are the bar path and cross paths insertion losses, respectively. For the paths that contain a leaf node, such as edge c, d, and b in Fig. 4(a), their insertion losses can be calculated with the following equation:

$$IL_{edge} = \sum IL_{PUC} \quad (4)$$

where $IL_{PUC}$ is the insertion loss of all the PUCs in the path, which corresponds to line 13. On the other hand, for the insertion losses of a branch (path) that contains a child node (lines 10 – 11), such as path a leading to the *TC_2* subtree in Fig. 4(a), the paths' insertion loss of *TC_1* can be calculated with Eq. (5). In the latter, $K_q$ represents the coupling factor of the child node *q* (in this case *TC_2*), $IL_{q,cross}$ is the insertion loss of child node *q* in the cross path, and $IL_{path}$ is the insertion loss between the parent node and the child node [32].

$$IL_{cross(bar)} = 10 \times \log(K_q) + IL_{q,cross} + IL_{path} \quad (5)$$

In this demonstration, we showcase the multicasting capabilities of our programmable processor driven by our control software layer. Specifically, we have implemented a 1-by-26 on-chip multicasting circuit. Port 0, connected to a 5 dBm source, serves as input while the rest of ports in Fig 1(a) act as receiving nodes. The result has been normalized with respect to the input power and measured at a central wavelength of 1550 nm, see Fig. 4(b). Our experimental approach began with a 1-by-1 circuit and progressively expanded the number of output ports up to 26. For each configuration, we have recorded the powers at all output ports and have placed them as columns of the results matrix in Fig. 4(b), so that the first and last columns represent the outgoing power distribution for the 1-by-1 and 1-by-26 cases, respectively. The resulting visualization of output power distribution versus number of receiving nodes exhibits a horizontal funnel pattern and demonstrates how evenly the signal can be shared across the chip ports. The yellow area systematically expands from the left side to the right, while the color tone changes to orange. This trend indicates that with an increasing number of ports, the power allocated to each port decreases, up to a maximum of -22 dB in the 1-by-26 case. Notably, the power intensity across all configured output ports remains uniform. Fig. 5 (c) offers another perspective to the power spreading problem, here we demonstrate that the average power exhibits a logarithmic decrease as the number of ports increases. The deviation shows a gradual increase with the number of targets: The minimum deviation is 0.663 dB, and the maximum is 1.31 dB for 1-by-2 and 1-by-26 multicasting, respectively. This increase in deviation is a consequence of the coupling factor dependence on the cross and bar insertion losses of the outgoing paths as shown in Eq. (3). The latter entails that in order to have an accurate compensation of path loss difference we require a precise measurement of each PUC's insertion loss. However, for this work we have relied on an average loss value for all the PUCs, therefore, as more of these are added to the circuit, more deviation is expected. In addition to path loss difference errors, optical crosstalk originating from a PUC can also contribute to non-uniform output powers. In the chip platform used in this work, we have measured that the optical crosstalk of individual PUCs is always below -25 dB. Note that we could also employ the on-chip photodetectors and the closed feedback loop to dynamically adjust or fine tune the coupling factors to correct all aforementioned errors starting from the analytical seed provided by the presented algorithm[22].

**Table 2. Pseudo-code of multicasting application**

| | Algorithm: auto_multicast |
|---|---|
| | **Function auto_multicast** *(input_port, output_ports, proportion)* |
| 1 | *multicasting_path* ← ∅ |
| 2 | *splitter_ratio* ← ∅ |
| 3 | *interconnect_paths* ← ∅ |
| 4 | **find** *interconnect_paths*<br># get interconnect path for each output |
| 5 | *tunable_pucs* ← GET_TUNABLE_PUCs(*interconnect_paths*) |
| 6 | **sort** *tunable_pucs*     # from child nodes to parent nodes |
| 7 | *tunable_pucs_properties* ← *{}* |
| 8 | **for** *tunable_puc* in *tunable_pucs* |
| 9 |     **for** *path* in GET_PATH (*tunable_puc*) |
| 10 |         **if** EXIST_CHILD_NODE (*path*) **is True** |
| 11 |             *ILs* ← **Eq. (5)** |
| 12 |         **else** |
| 13 |             *ILs* ← **Eq. (4)** |
| 14 |     **update** *tunable_pucs_properties* **with** *ILs* |
| 15 | **calculate** *k* **with Eq. (3)**<br># using values of *proportion* and *ILs* from *tunable_pucs_properties* |
| 16 | **add** *k* **into** *splitting_ratio* |
| 17 | **update** *multicasting_path*<br># using values of *splitting_ratio* and *interconnect_paths* |

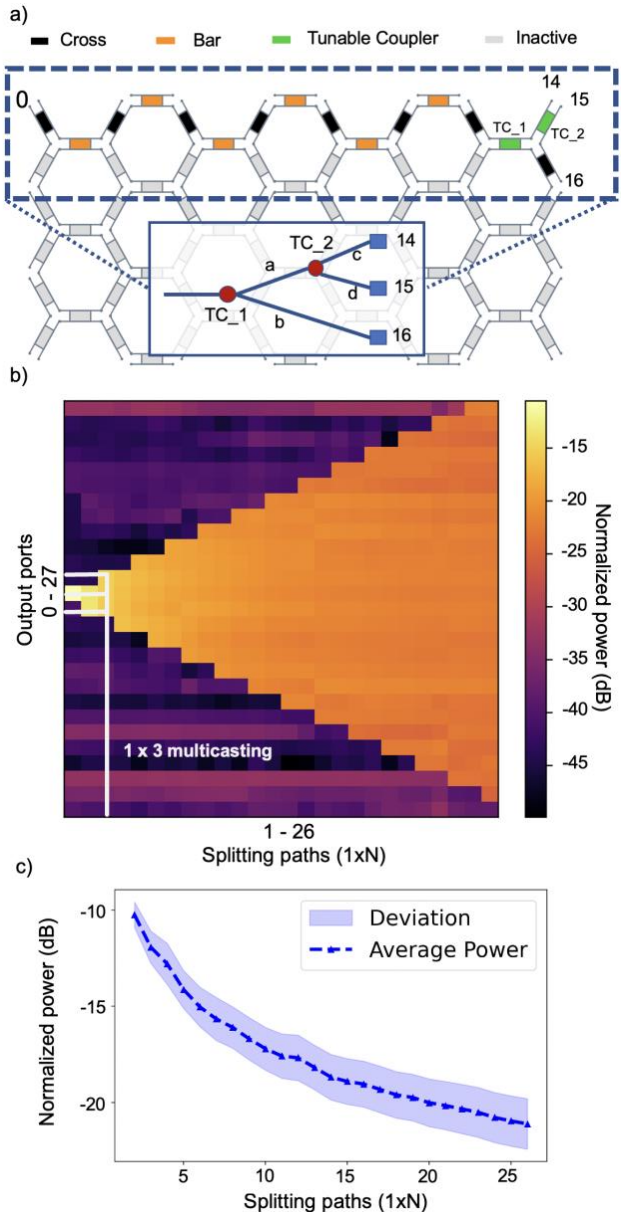

Fig. 4. Optical multicasting demonstration: a) Programmable processor synthesizing a 1x3 multicasting configuration. The inset shows the detailed graphs b) Output power colormap as a function of the splitting paths starting from 2 (second leftmost column) and 26 (rightmost column). The white lines point to the measured output powers for the configuration in a).

## 4. DISCUSSION AND CONCLUSION

The results presented in this manuscript experimentally demonstrate optical networking applications on a PIP platform, with a software layer capable of adapting to dynamic network conditions and ensuring optimum performance. In comparison with the state-of-the-art, our work represents a pioneering achievement by introducing flexible software for convenient control and reconfiguration of a complex photonic processor. Our experiments have been conducted on a large mesh of 72 PUCs with path-finding time of less than a half millisecond, compared to previous results in this field with lower size meshes [23]. This substantiates the scalability of programmable integrated circuit and establish a significant milestone for future developments in software-defined photonic integrated circuit advancements. Moreover, we implement and experimentally validate for the first-time optical circuit switching into a general-purpose processor with -25 dB of crosstalk. This represents a significant advancement in the field where only simulation results were reported for switching [23] [25] [26]. To improve crosstalk, forthcoming work in the calibration and characterization process should be carried out to enhance the accuracy and tunning of the PUCs. In addition, better splitters and PUC architectures have been demonstrated to obtain crosstalk values around -50 dB [33], [34]. Another critical figure of merit is the accumulated loss. In this sense, both better chip-fiber coupling mechanisms [35] and lower-loss PUCs [36] have been demonstrated, enabling in a short-term future less than 3 dB total system loss with a similar scale as the presented in this work. Finally, we experimentally implement a multicasting configuration of up to 1x26 signals with maximum deviation of 1.31 dB for all output powers, compared to the general-purpose literature, where multicasting is only reported theoretically [32].

In summary, this work showcases networking functions using a programmable photonic processor and employing a software layer based on graph theory. Our implementation facilitates the automatic configuration of diverse applications on programmable PICs with remarkable efficiency, reconfiguration speed and flexibility. We experimentally demonstrated an optical interconnect, optical switching and optical multicasting, everything running in the same photonic hardware. Overall, the results provided in this work address some fundamental limitations in optical networks by using integrated photonic solutions, leading the way to next-generation optical devices in datacenter applications.

**Funding Information.** This work has received funding from the European Union's EIC Transition programme under grant agreement No. 101057934 (INSPIRE Project), ERC Starting Grant programme under grant agreement No. 101076175 (LS-Photonics Project) and Horizon Europe research and innovation programme under grant agreement No. 101092766 (ALLEGRO Project)

**Acknowledgment**. We thank the software team of iPronics, Programmable Photonics S.L. for their technical support. DSJ and ZX are pursuing an Industrial PhD in Telecommunications at Universitat Politècncia de València (UPV).

**Zhenyun Xie** obtained his M.Sc degrees in telecommunication engineering at Universitat Politècnica de València, Valencia, Spain (UPV), and received the best master thesis award from Cátedra DEXTROMÉDICA in 2022. Currently, he is software engineer at iPronics, Programmable Photonics S.L., Valencia, Spain. Meanwhile, he is working toward a Ph.D. in telecommunication at UPV. His research interests focus on advanced programming methodologies for photonic processors.

**David Sánchez Jácome** obtained in 2016 his BSc in Electrical Engineering from University San Francisco de Quito (USFQ), Ecuador. Part of these studies were undertaken at the University of Ottawa, Canada. In 2019 he finished his master's studies in the Europhotonics consortium obtaining the degrees of MSc in Applied Physics from the University of Aix-Marseille (AMU), France and MSc in Optics and Photonics from the Karlsruhe Institute of Technology (KIT), Germany. He currently works as software and application lead in iPronics where he is also enrolled as an industrial PhD candidate in collaboration with the Photonics Research Lab (PRL) of the Universitat Politècnica de València (UPV). His work focuses on developing advanced routines and applications for integrated software-defined photonic technologies.

**Dr. Luis Torrijos Morán** holds a B.S. degree in telecommunication and a M.S. degree in electronics engineering from the Universitat de València. In 2021 he graduated with honors in his Ph.D from the Universitat Politècnica de València (UPV) and received the Extraordinary Ph.D. Thesis Award from the same university. He made a research stay with the Photonics Research Group in Ghent University, Belgium and continued with a two-year postdoc with the Photonics Research Labs (PRL) in photonic integrated circuits. Currently, he serves as a photonics engineer in iPronics, focusing on the development of next-generation programmable photonic systems. His academic contributions include the authorship of over ten articles published in high-impact scientific journals, and active participation in more than fifteen congress papers.

**Dr. Daniel Pérez-López** (born 1991) is a scientist and deep-tech entrepreneur in the field of programmable integrated photonics, renowned for his pioneering contributions to research, development, and commercialization of programmable integrated photonic circuits. He holds a PhD in Telecommunications with a focus on integrated microwave photonic processors from the Photonics Research Labs at Universitat Politècnica de València (UPV). His educational journey includes a Postdoc, Master's and Bachelor's in Telecommunication Technologies and industrial expertise (UPV, Optoelectronics Research Centre). With his team at iPronics, he put in the market the world-first software defined general purpose photonic processor. As Co-founder and CTO, he now leads pivotal research advances in hardware architectures, components, and software for practical large-scale systems for flexible optical networking and processing.